\documentclass{epl}

\usepackage{graphicx}
\usepackage{bm}
\usepackage{amsmath}
\usepackage{amssymb}
\usepackage{amsfonts}

\title{Giant orbital moments are responsible for the anisotropic magnetoresistance of atomic contacts}
\author{G. Aut\`es\inst{1}, C. Barreteau\inst{1}, M.C. Desjonqu\`eres\inst{1},  D. Spanjaard\inst{2}, \and  M. Viret\inst{3}}

\institute{                    
  \inst{1} CEA Saclay, DSM/IRAMIS/SPCSI, B\^atiment 462, F-91191 Gif sur Yvette, France\\
  \inst{2} Laboratoire de Physique des Solides, Universit\'e Paris Sud,   B\^atiment 510, F-91405 Orsay, France  \\
  \inst{3} CEA Saclay, DSM/IRAMIS/SPEC, B\^atiment 772, Orme des Merisiers, F-91191 Gif sur Yvette, France 
}


\begin{document}

\maketitle

\begin{abstract}
 We study here, both experimentally and theoretically, the anisotropy of magnetoresistance 
 in atomic contacts. Our measurements on iron break junctions reveal an abrupt and hysteretic 
 switch between two conductance levels when a large applied field is continuously rotated.
 We show that this behaviour stems from the coexistence of two metastable electronic states 
 which result from the anisotropy of electronic interactions responsible for the enhancement of orbital  
magnetization. In both states giant orbital moments appear on the low coordinated
central atom in a realistic contact geometry. However they differ by their orientation,
parallel or perpendicular, with respect to the axis of the contact.
 Our explanation is totally at variance with the usual model based on the band 
 structure of a monatomic linear chain, which we argue cannot be applied to 3d ferromagnetic metals.  
\end{abstract}

The electronic transport in magnetic systems with reduced dimensionality has been
intensively studied during the last decade. Indeed, beyond the exciting physics
which is being unveiled, this research area is also of strategic importance
for technological applications in spintronics such as spin valves,
magnetic memories and read heads.
In this context, nanocontacts made by mechanically or electrically
controllable break junctions, electro-deposition or using the tip of
a scanning tunneling microscope have attracted a great deal of interest.
In the ballistic regime when the contact dimensions are below
the conduction electron mean free path, Ohm's law breaks down
and new physical properties emerge. When the contact diameter
becomes comparable to the Fermi wavelength of the conduction electrons,
quantum effects appear. For instance, an infinite magnetic monatomic wire 
has a conductance quantized in units of $G_0=e^2/h$ (for very small
bias voltages), the number of quanta being given by the number of conduction channels,
{\sl i.e.}, the number of linearly independent wave functions at the
Fermi level of positive wave vector \cite{Fisher1981}.
Furthermore, the electronic structure depends on the direction of magnetization due
to spin-orbit coupling. Consequently, a sizeable anisotropic orbital
component of the magnetization appears and the number of bands at
the Fermi level may change.
This phenomenon is called ballistic anisotropic magnetoresistance (BAMR)
or atomic anisotropic magnetoresistance (AAMR), and it has been shown to
be much larger than the AMR observed in bulk samples.

\begin{figure}[!fht]
\begin{center}
\includegraphics*[width=8cm]{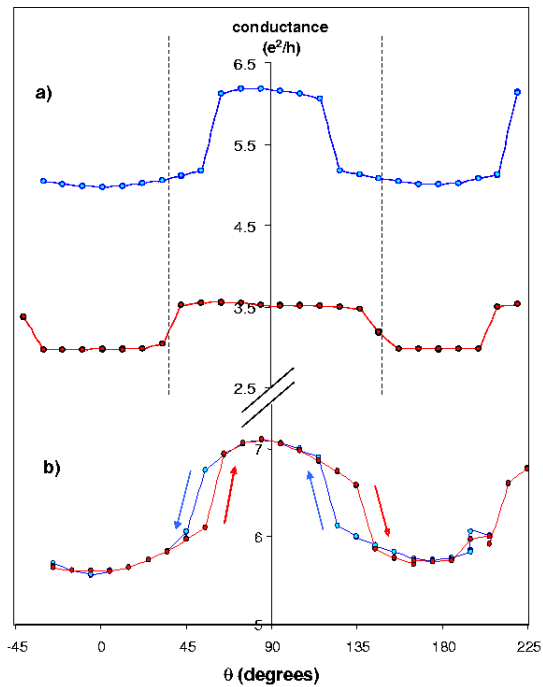}
\caption{Variation of conductance with the direction $\theta$ of the applied magnetic field 
(see Fig. \ref{fig:geom}) for different Fe atomic contacts.
 a) The switching angles between the two resistive states can be different
depending on the specific atomic arrangements.
b) Some hysteresis appear when rotating the field back and forth.}{\label{fig:expAMR}}
\end{center}
\end{figure}

A series of experiments using the break junction technique has
been carried out on 3d magnetic materials (Fe, Co, Ni). Narrow
constrictions are slowly broken by mechanically bending the
substrate and the conductance typically falls to zero passing
through different discrete levels. As breaking is a highly non
monotonic process, each level has been attributed to the
conductance of a different stable atomic geometry at the contact
\cite{Scheer98}. Clear conductance quantization is routinely
 observed in Au, but 3d metals behave differently \cite{Untiedt2004}.
It is actually remarkable that almost any value of conductance can
be stabilized in a breaking experiment (with a bit of patience).
Some contacts can also be stable for hours, or even days. This is
essential for magnetoresistance measurements \cite{Viret2002} where
a large external field needs to be slowly swept or rotated while
monitoring the resistance. A typical set of measurements on Fe
under a 2T rotating magnetic field at 4.2K is shown in
Fig.\ref{fig:expAMR}. These results are consistent with previous
reports showing a large AMR effect in atomic sized contacts as
well as the, often observed, two level behaviour
\cite{GabureacThesis,Viret2006,Bolotin2006}. Two supplementary
observations can be made from Fig.\ref{fig:expAMR}:
 i) the two conductance levels can have a different domain of angular stability
({\sl i.e.}, the angle at which the transition occurs is not always $45^\circ$,
ii) some hysteresis is present when rotating the external field back and forth.
 These characteristics have not yet been addressed and a quantitative
explanation for the AMR in a realistic atomic geometry is still lacking.

A first qualitative explanation of the step-like variation of the conductance with the direction of
the applied magnetic field relies on the simple geometry in which the
contact is modelled by a linear chain of atoms. Ab-initio calculations \cite{Viret2006,Velev2005} 
for ferromagnetic 3d transition metals have shown that the degenerate weakly dispersive $\delta$ bands 
arising from the coupling of nearest neighbour $3d_{xy}$, on the one hand, and $3d_{x^2-y^2}$,
on the other hand, atomic orbitals are split by the spin-orbit interaction
except when the magnetization is perpendicular to the wire.
As a consequence when the Fermi
level lies close to the edges of the $\delta$ bands, one of the bands may
cross the Fermi level when the field is rotated thus modifying the number of conduction
channels and producing an abrupt variation of
the conductance of amplitude $e^2/h$.
Similar arguments have been put forward by Sokolov et al. \cite{Sokolov2007}.

Although quite attractive for its pedagogical character, this interpretation presents several weak points.
Mainly it is known from experimental and theoretical works
that only 5d elements (Ir, Pt, Au) have the ability of
forming true monatomic wires several atoms long \cite{Smit2001,Bahn2001,Smit2003,Fernandez2007}, while 3d elements
(Fe, Co, Ni, Cu) can only form very short suspended chains of
at most three atoms \cite{Rodrigues2003,Sato2006,Amorim2007}. 
Therefore, there is a need for a more realistic description of the constriction
region connected to the electrodes.

In a realistic atomic geometry, the electronic structure cannot be described in terms
of bands due to the loss of translational symmetry in the contact region.
Consequently  the conductance is no longer quantized in units of $G_0$.
Furthermore in the 1D theoretical model the removal of degeneracy due to spin-orbit 
coupling plays a central role in determining the band structure.
However, in the works quoted above \cite{Velev2005,Viret2006,Sokolov2007}, the Local Spin Density
Approximation (LSDA) or Stoner-like tight-binding models were used which
underestimates the orbital moment, {\sl i.e.}, the orbital polarization effects.
Indeed, in these approaches, intra-atomic electronic interactions between electrons
in different orbitals are averaged,  thus their orbital dependence is neglected.
Desjonqu\`eres {\sl et al.} \cite{Desjonqueres2007a,Desjonqueres2007b} have investigated these effects in details by using
an Hartree-Fock (HF) Hamiltonian including all matrix elements of the Coulomb electronic interactions
with their full orbital dependence.
They showed that in monatomic chains there is a strong enhancement of the 
orbital magnetic moment and of its anisotropy without
changing significantly the spin moment.
In addition  the splitting of the minority spin $\delta$ bands at $\theta=0$ 
is greatly affected by the value of the
orbital moment. The magnetic anisotropy is considerably increased and may locally
reach several meV per atom. Therefore, it is likely that the external magnetic 
field will not be large enough to align all the magnetic moments
and one can expect a non-collinear magnetic configuration in the constriction region.
For all these reasons, orbital polarization effects cannot be ignored
in the calculation of the conductance in 1D models as well as for more complex
geometries with atoms in a low coordination configuration, and non collinearity of
spins has to be allowed.

An improved modelling of the break junction experiment should therefore
consist of three steps: i) the choice of more realistic structural arrangements,
 ii) the accurate computation of the electronic and magnetic
structures of the system,  iii) finally, the calculation of the ballistic electronic transport
through the junction.

\begin{figure}[!fht]
\begin{center}
\includegraphics*[width=7cm]{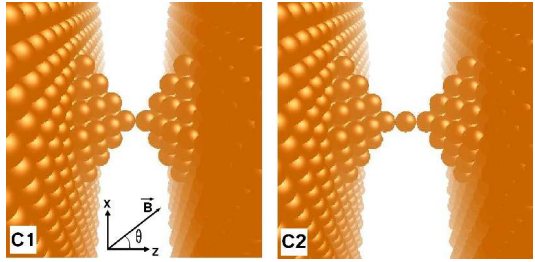}
\caption{Atomic configurations C1 and C2 of the break junction. The $x$ axis
is parallel to the sides of the basis of the square pyramids.}{\label{fig:geom}}
\end{center}
\end{figure}

In order to evaluate the influence of local geometrical
arrangements we have considered  two different atomic configurations C1 and C2 (Fig. \ref{fig:geom})
for an iron break junction. For both geometries the leads are made of two semi-infinite $(001)$ bcc surfaces
while the atomic contact region in the gap separating the leads is built from two
head-to-head square pyramids without (geometry C1) and with (geometry C2)
an extra atom between the two apices. All atoms are separated by the
bulk nearest neighbour spacing.
These two model geometries are plausible structures as
shown by recent high resolution transmission electron microscopy (HRTEM)
imaging \cite{Sato2006} and from realistic molecular dynamics simulations \cite{Amorim2007}
of copper nanowires under stress.

For the second step we have used
a tight-binding method with a $s$, $p$ and $d$ atomic orbital basis set
that provides an accurate description
of the magnetic properties of iron in various geometries and dimensionalities \cite{Autes2006}.
Spin-orbit coupling and the Zeeman energy  term are included and non-collinear magnetic configurations
are also allowed.  The electronic interactions are treated in two different ways: a
simple scheme using a Stoner model and  a second one (HF) taking into
account the orbital dependence of electron-electron interactions in a full Hartree-Fock
scheme \cite{Desjonqueres2007a,Desjonqueres2007b}. The technical details and the values
of the parameters can be found in Ref. \cite{Desjonqueres2007b}.

Finally, in the last step, the self-consistent Hamiltonian derived from
the electronic structure calculation is used to compute
the ballistic electronic conductance through the atomic
contact. 
This involves the calculation of the trace of the transmission matrix \cite{Fisher1981,Dattabooks}:

\begin{equation}
 T(E)=Tr(\Gamma_L G_C \Gamma_R G_C^\dag),
\end{equation}

\noindent where $\Gamma_{\text{L(R)}} $ is the imaginary part of the self energy of the left (right)
electrodes, and $G_C$ the Green function of the scattering region, and $G_C^\dag$ it's hermitian conjugate.

\begin{figure}[t]
\begin{center}
\includegraphics*[width=\linewidth]{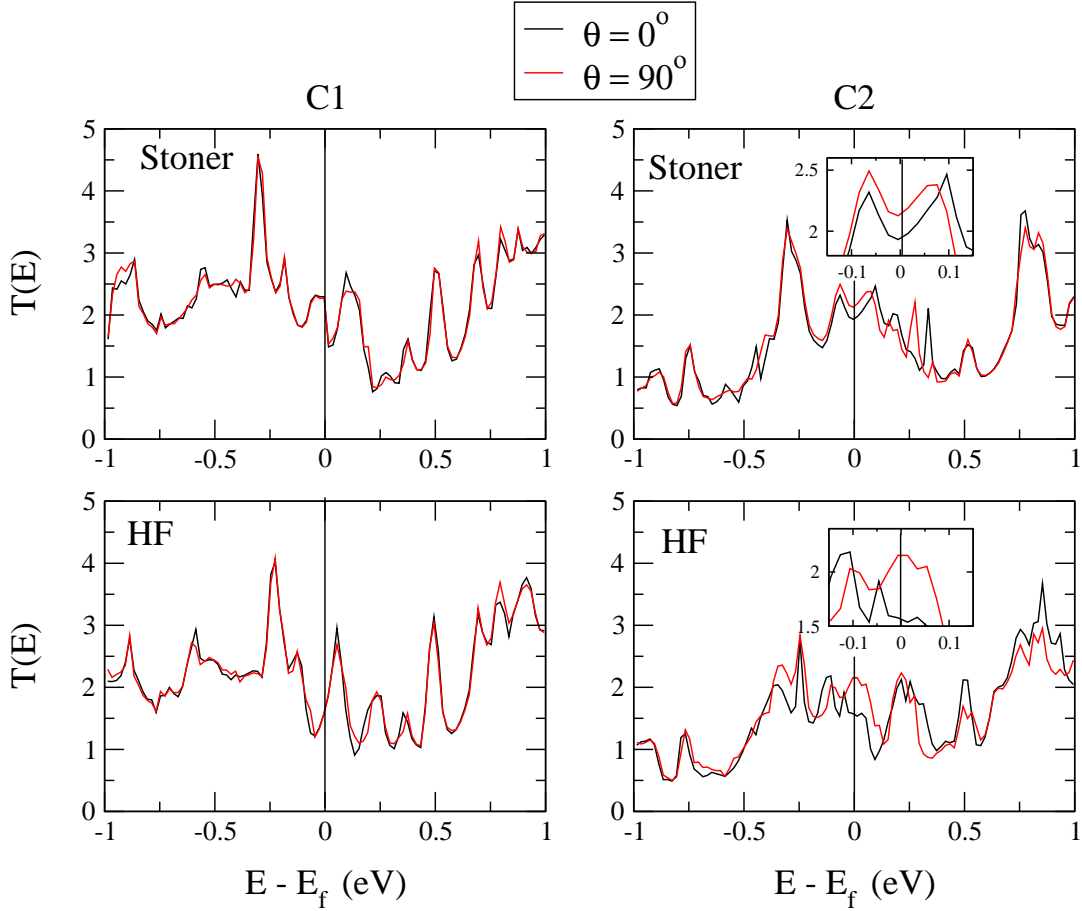}
\caption{Transmission coefficient $T(E)$ as a function of energy for $\theta =0^\circ$
(black curve) and $\theta =90^\circ$ (red curve) calculated
for the two systems C1 and C2 (See Fig. \ref{fig:geom}) with the Stoner and the HF  electronic
interaction models. An enlarged view of
 $T(E)$ around the Fermi level is shown in the insets of the figures
for the C2 system.}{\label{fig:TE}}
\end{center}
\end{figure}

In order to investigate possible magnetoresistive
effects, we have computed the electronic and magnetic structures of
the systems C1 and C2 for an applied magnetic field lying in the planes limiting
the electrodes or perpendicular to them (along the pyramid axis), namely
along the $x$ or $z$ axis (see Fig. \ref{fig:geom}).
The transmission coefficient as a function of the energy $E$ is then
computed for the two electronic interaction models. The results, shown in Fig \ref{fig:TE}, underline the importance
 of the contact geometry. It comes out that the anisotropy of system C1 (left panel of Fig. \ref{fig:TE}) is
extremely small and one cannot expect any significant AMR
for such a geometry. On the contrary, the system C2 (right panel of Fig. \ref{fig:TE})
shows an anisotropic transmission coefficient due to the monatomic wire configuration of the central
 atom which induces a stronger magnetic anisotropy. Hence, the presence of a short wire between the 
 electrodes is necessary to induce a large AMR in atomic contacts. As expected the transmission coefficients are no longer integers because
 the complex geometry of the system gives rise to elastic scattering at its boundaries.
Finally, the overall magnetic anisotropy of the transmission coefficient curve $T(E)$
for the C2 geometry is stronger with the HF model than with the Stoner one.
However it is interesting to note that around the Fermi level
the anisotropy obtained with the Stoner model is noticeable 
(see inset of right panel of Fig. \ref{fig:TE}).

\begin{figure}[t]
\begin{center}
\includegraphics*[width=7cm]{figure4.eps}

\includegraphics*[width=7cm]{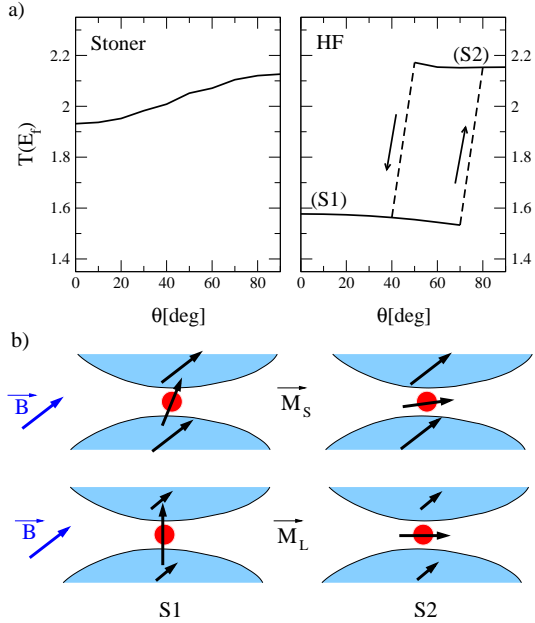}
\caption{
a) Transmission coefficient at the Fermi level $T(E_f)$ as a function of the magnetization direction $\theta$   
for the atomic configuration C2 with the Stoner (left) and HF (right) electronic interaction models.
For the HF model, S1 and S2 denote 
the two electronic solutions (see Text) schematically depicted in b).
The arrows indicate the direction of the spin ($\vec{M_S}$) and orbital ($\vec{M_L}$) local magnetic moments
in the pyramids (in blue) and at the central atom (red circle) of the atomic configuration C2.
Their length is roughly proportional to the value of the moment. Note that
the spin and orbital moments in the leads are aligned along the applied magnetic field $\vec{B}$,
and that there is a range of angles in which both solutions exist.}{\label{fig:Ttheta_and_solution}}
\end{center}
\end{figure}

We now concentrate on the C2 geometry since it is the one leading to significant anisotropic effects. In order
to compare the two electronic interaction models, we have computed the conductance
(in units of $e^2/h$) given by the transmission coefficient at the Fermi
level $T(E_f)$ of the system C2 with a magnetization direction $\theta$ varying from $0^\circ$ to $90^\circ$
(Fig. \ref{fig:Ttheta_and_solution} a). The calculated absolute values of AMR ($10.2\%$ with Stoner and $36.6\%$ with HF)
are in the same range as those found in experiments.
The results show that in the Stoner model the conductance varies continuously following a sine-like curve.
On the contrary, the HF model reproduces the measured step-like curve with an abrupt change of the conductance appearing between
$\theta=40^\circ$ and $\theta=70^\circ$. 
With this model the step-like behaviour can be attributed to the switch between two metastable electronic
solutions, S1 and S2 described schematically in Fig.\ref{fig:Ttheta_and_solution}b. 
These two states differ by the local spin and orbital magnetic configuration on the three
central atoms,whereas the magnetization of all other atoms is almost  aligned along the magnetic field.
More precisely for S1, the spin moment on the central atom is canted towards the axis of the wire with respect
to the magnetic field orientation while for S2 it is canted towards the perpendicular to the wire
(note that in the Stoner model all the spins are nearly collinear due to a smaller anisotropy).
The distinction between S1 and S2 is even more pronounced on the local orbital moment.
Indeed for S1 the orbital moment of the central atom is dramatically enhanced 
(2.5$\mu_B$) and almost aligned along the axis of the wire.
For the S2 solution it is smaller (1.5$\mu_B$) and almost perpendicular to this axis.
Actually the orbital moment of the central atom is nearly independent on $\theta$ in the respective
domains of stability of S1 and S2.
 The change of the magnetic field direction is therefore responsible  for the switch between
S1 and S2 and a complete rearrangement of the electronic structure  takes place for an angle
$\theta$ between $40^\circ$ and $70^\circ$.
This effect induces the abrupt change in the conductance. Let us note that the angle $\theta$ at which the switch
occurs is not precisely defined and might be different when $\theta$ increases or decreases leading to 
 hysteresis effects as observed experimentally.
The possibility of  multiple solutions is intimately linked to the HF model that is known to generate several
metastable states in low dimensional systems, as already observed in our previous studies
\cite{Desjonqueres2007a,Desjonqueres2007b}.

In conclusion we have measured the magnetoresistance of an iron break junction as a function of the
direction of the applied magnetic field, and observed an abrupt variation for an angle
around $45^\circ$ with an hysteresis effect. This behaviour has been explained
by a calculation of the transmission coefficient on a realistic geometry and an
elaborated model of the electronic structure, {\sl i.e.}, including not only
the spin-orbit coupling but also the orbital polarization effects as well
as the Zeeman energy term and allowing a non collinearity of spin moments.
We find that the stepwise variation of the magnetoresistance is due to
the existence of two metastable electronic states which differ mainly by the direction
of the spin and orbital moment on the central atom and may explain 
the  observation of an hysteresis when the magnetic field is rotated
back and forth. Orbital polarization effects,
which are known to strongly enhance the magnetic anisotropy energy in low
dimensional systems, are of prime importance for this interpretation since
a Stoner-like Hamiltonian fails to reproduce the experimental results.
Therefore, the large AMR effects observed  do not originate from 
the ballistic nature of transport
but rather by the low dimensionality of the atomic contacts.

\end{document}